\documentclass{iopconfser}
\usepackage[utf8]{inputenc}
\usepackage{amsmath} % For math environments
\usepackage{graphicx}
\usepackage[dvipdfmx]{color}
\usepackage{bm}

\begin{document}

\title{Revisiting $\mu$SR Studies of Ion Dynamics in the Light of Extended Kubo-Toyabe  Model}

\author{Takashi U. Ito$^{1}$ and Ryosuke Kadono$^{2}$}

\affil{$^1$Advanced Science Research Center, JAEA, Tokai, Ibaraki 319-1195, Japan}
%\vspace{2mm}
\affil{$^2$Institute of Materials Structure Science, KEK, Tsukuba, Ibaraki 305-0801, Japan}

\email{ito.takashi15@jaea.go.jp, ryosuke.kadono@kek.jp}
%\email{ito.takashi15@jaea.go.jp}

\begin{abstract}
The dynamical Kubo-Toyabe (dKT) function is extended to describe the spin relaxation under the coexisting dynamical and static internal magnetic fields.  A detailed re-evaluation of the previous $\mu^\pm$SR data in Na$_x$CoO$_2$ using this function disfavors the conventional interpretation based on sodium-ion diffusion and instead supports the $\mu^+$ self-diffusion scenario. This also resolves the long-standing inconsistencies in the dKT-function-based $\mu$SR studies on ion diffusion from the viewpoint of classical over-barrier-jump mechanism.
\end{abstract}
\vspace{-5mm}

%\section{Introduction}

Muon spin rotation/relaxation ($\mu$SR) has been regarded as a powerful local probe for investigating ion dynamics in various non-magnetic materials including those for batteries. The fluctuation frequency $\nu$ of random local fields from nuclear magnetic moments deduced by analyzing $\mu$SR spectra using the dynamical Kubo-Toyabe (dKT) function is interpreted as the jump frequency of ions surrounding muon \cite{Sugiyama:09,Sugiyama:11,Sugiyama:12,Mansson:13,Baker:11,Umegaki:17,Sugiyama:20,Ohishi:22a,Ohishi:22b,Umegaki:22,Ohishi:23,Forslund:25}. This interpretation relies on the assumption that the muon itself remains immobile in the time scale of $\mu$SR measurements ($\sim$10$^1$ $\mu$s), so that the cause of the fluctuation is solely attributed to the motion of surrounding ions carrying nuclear magnetic moments.

However, this conventional approach entails a number of issues to be resolved. Most notably, in layered cobalt oxides ($A_x$CoO$_2$, $A=$ Li, Na ) used as cathode materials in batteries, muons experience magnetic fields from both diffusing ions ($^{6,7}$Li$^+$, $^{23}$Na$^+$) and immobile ions ($^{59}$Co) \cite{Sugiyama:09,Mansson:13,Sugiyama:20,Ohishi:22b,Ohishi:23}. Such situations do not satisfy the important prerequisites of the strong collision model incorporated into the dKT function for describing the fluctuation of internal field ${\bm B}$, so that the justification for using the dKT function in the analysis of these compounds is not obvious.

%\section{The Extended Dynamical Kubo-Toyabe Model}

To address this issue, we have developed an extended dKT function, $G_z^{\rm EA}(t;Q,\Delta,\nu)$, based on a model that considers ${\bm B}$ as a vector sum of static and dynamical (fluctuating) components \cite{Ito:24,edKT}. The model is characterized by an Edwards-Anderson-type autocorrelation function \cite{Edwards:75,Edwards:76}, 
\begin{equation}
C(t)=\gamma_\mu^2 \langle B_\alpha(t)B_\alpha(0)\rangle/\Delta^2=(1-Q) +Q\exp(-\nu t), \:\:(\alpha=x,y,z)
\end{equation}
where $Q= \Delta_{\rm d}^2 / (\Delta_{\rm s}^2 + \Delta_{\rm d}^2)$ represents the relative weight of the fluctuating component with $\Delta_{\rm s}^2 + \Delta_{\rm d}^2 =\Delta^2$, $1-Q $ is that of the static component, and $\nu$ is the fluctuation frequency. The conventional dKT function corresponds to the specific case of $Q=1$. (The above definition of $Q$ corresponds to $1-q$ in Ref.~\cite{Edwards:75,Edwards:76}.) 

\begin{figure}[t]
\begin{center}
  \includegraphics[scale=0.5,clip]{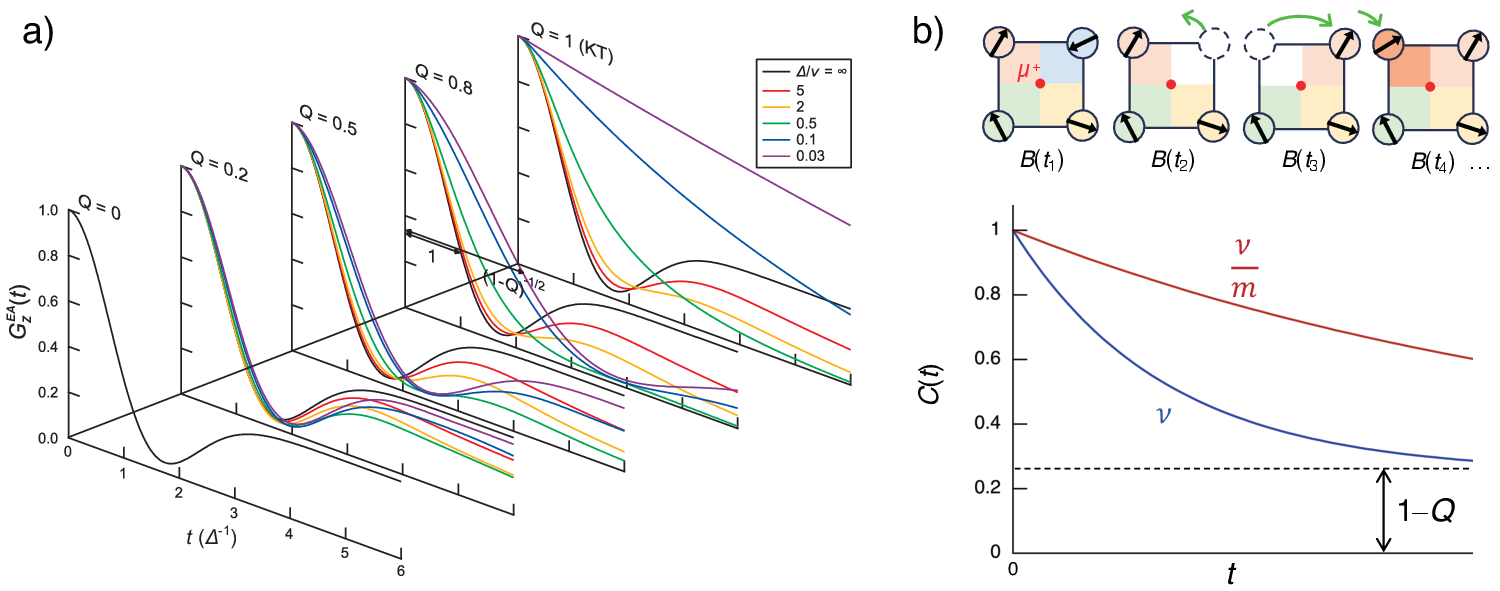}
 \caption{ (a) Typical examples of the extended dKT function $G_z^{\rm EA}(t;Q,\Delta,\nu)$ at zero field with varying $Q$, where curves are given for $\Delta/\nu=0.03$, 0.1, 0.5, 2, 5, and infinity. The decrease in the effective linewidth from $\Delta$ to $\sqrt{1-Q}\Delta$ with $\Delta/\nu\rightarrow0$ is clearly seen at $Q=0.8$. (b) Autocorrelation function $C(t)$ in ion diffusion, where $1-Q$ is the static fraction associated with immobile ions. When it takes $m$ steps (accompanying field fluctuations at each step) for replacing all ions around the muon, $C(t)$ decays with a reduced rate $\nu/m$.  } 
  \label{fig1}
 \end{center}
\end{figure}

As shown in Figure \ref{fig1}(a), a distinct feature of $G_z^{\rm EA}(t)$ is the large difference in the $\nu$ dependence of the lineshape between $Q<1$ and $Q=1$. Unlike the conventional dKT function ($Q=1$), the motional narrowing is strongly suppressed for $Q<1$ even in the limit of fast fluctuations ($\nu\gg\Delta$). Instead, the overall lineshape is always dominated by the static component, asymptotically approaching a static KT function characterized by the reduced linewidth $\sqrt{1-Q}\Delta$.
Here, it should be stressed that the $\nu$ in eq.~(1) does not necessarily correspond to the ion jump frequency.  As shown in Figure \ref{fig1}(b), provided that $m$  ($>1$) is the number of steps required for replacing all ions around the muon to reset ${\bm B}(t)$ completely, the effective decay rate of $C(t)$ is $\nu/m$, and  the fluctuation rate obtained in the analysis based on eq.~(1) is underestimated by the factor $1/m$.

%\section{Origin of spin relaxation in Na$_{0.7}$CoO$_2$}

In view of this newly developed relaxation function, the cathode material Na$_{x}$CoO$_2$ provides an excellent test case \cite{Mansson:13, Sugiyama:20}, as both $\mu^-$SR and $\mu^+$SR data are available, offering the opportunity to examine the consistency of interpretation derived from these data at different probe sites \cite{Ito:25b}.  In $\mu^-$SR experiment, the probe is the polarized muonic O atom situated at a regular O site. Here, the static field from the surrounding $^{59}$Co nuclei is overwhelmingly
dominant ($\sim$97\% of entire $\Delta$ according to
Ref.\cite{Sugiyama:20}), corresponding to a very small dynamical fraction, $Q \approx 0.06$.
For such $Q$ value, the $G_z^{\rm EA}(t)$ function predicts that the $\mu$SR spectra should appear almost entirely static, regardless of the Na-ion jump rate.  This is in excellent agreement with the $\mu^-$SR data, which show virtually no variation with temperature. This validates our model function and confirms that the $^{59}$Co nuclear moments are indeed static on the $\mu$SR time scale.

\begin{figure}[t]
\begin{center}
 \includegraphics[scale=0.5,clip]{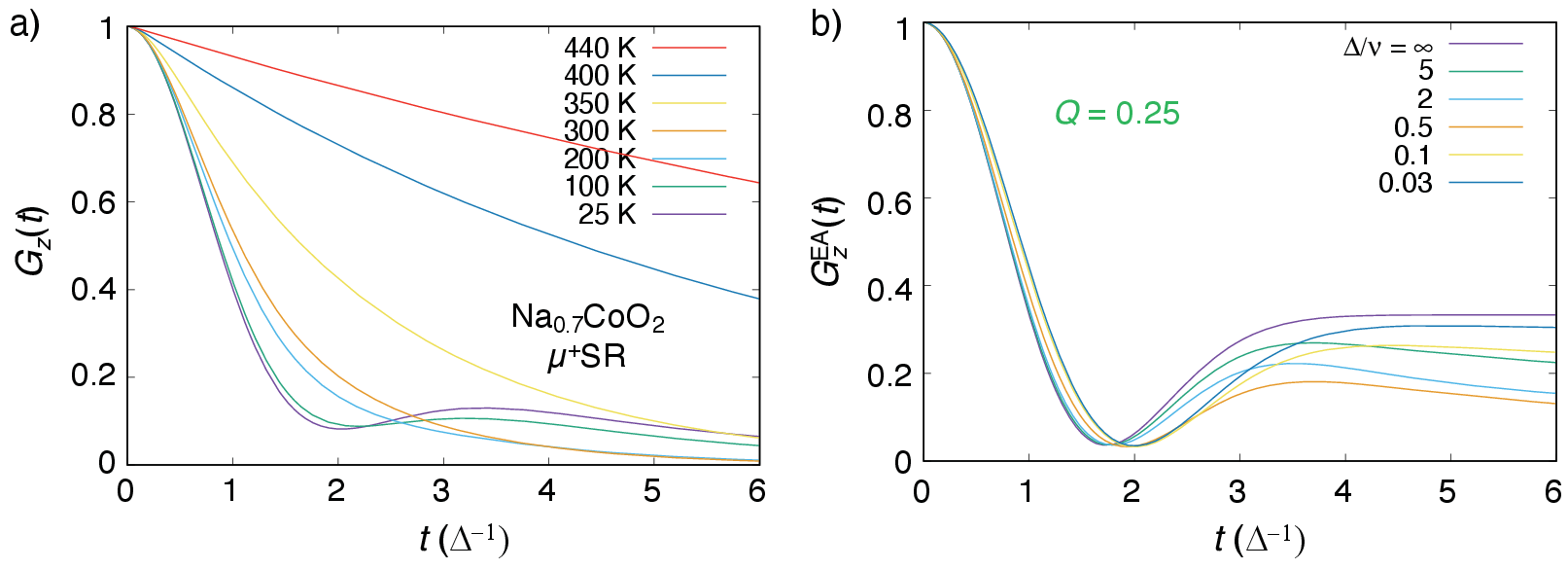}
 \caption{(a) ZF-$\mu^+$SR spectra in Na$_{0.7}$CoO$_2$ reproduced using the data reported in Ref.\cite{Mansson:13}. (b) The extended dKT function for $Q=0.25$ corresponding to the case of Na$_{0.7}$CoO$_2$.} 
 \label{fig2}
 \end{center}
\end{figure}

In contrast, the $\mu^+$SR data show significant motional narrowing at
high temperatures (see Figure \ref{fig2}(a)). For the interstitial $\mu^+$, the static field
from $^{59}$Co is still dominant ($\sim$87\% of entire $\Delta$
according to Ref.\cite{Sugiyama:20}), corresponding to a dynamical
fraction $Q\approx 0.25$. However, as shown in Figure \ref{fig2}(b), the lineshape of $G_z^{\rm EA}(t)$ with $Q=0.25$ is close to that of a static KT-like with little sensitivity to $\nu$. This prediction is in stark qualitative contradiction with the behavior seen in the experimental data which is well represented by the conventional dKT function ($Q=1$). 

This inconsistency also suggests that the similarity in the temperature dependence of $\nu$ between $\mu^+$SR and $\mu^-$SR results below $\sim$300 K (which is argued as evidence for the sodium diffusion) is accidental, apart from the large difference at higher temperatures \cite{Sugiyama:20}. 

Thus the only way to attain a coherent interpretation of $\mu^\pm$SR results is to depart from the long-held picture of ``immobile muons probing Na-ion diffusion.'' To be consistent with the extended dKT model, the observed motional narrowing in $\mu^+$SR can only be explained when the muon itself is undergoing diffusive motion relative to the static Co nuclei.

%\section{Cation molecule dynamics in hybrid perovskites}
Ion diffusion is not the only case in solids where static internal magnetic fields microscopcially coexist with fluctuating ones at muon sites. Hybrid lead iodide perovskites, $A$PbI$_3$ with the $A$ site consisting of organic cation molecules (e.g., CH$_3$NH$_3$) are one such example in which the fields consists of those from the organic molecules and PbI$_3$ lattice with the former fluctuating due to the random rotational motion. The $\mu$SR spectra measured in these materials at low temperatures are seemingly described by the quasi-static dKT function, but they maintain the Gaussian lineshape while evolving with increasing temperature (similar to the case of $Q=0.8$ in Fig.~\ref{fig1}(a)) in contrast to the exponential decay of the dKT function ($Q=1$). In fact, analyzing these spectra with the dKT function gives peculiar results: $\Delta$ decreases with increasing temperature, and the fluctuation frequency $\nu$ shows a strange peak instead of the expected monotonic increase \cite{Hiraishi:23}. On the other hand, $G_z^{\rm EA}(t)$ reproduces the temperature variation of the spectrum well with fixed $Q$ and $\Delta$, and the temperature variation of $\nu$ obtained by curve fitting shows the expected behavior for the thermal activation \cite{Ito:24}. From this example, we can conclude that the extended KT function can properly describe realistic ion dynamics.

%\section{Discussion}
One of the major issues associated with the interpretation of $\mu$SR data based on the dKT-function-based analysis is that the attempt frequency ($\nu_0$) derived from the temperature dependence of $\nu$ via the Arrhenius relation, $\nu=\nu_0\exp(-E_{\rm a}/kT)$, is  reported to be consistently in the order of $10^6$--$10^9$~s$^{-1}$ \cite{Sugiyama:09, Sugiyama:11,Sugiyama:12,Mansson:13, Baker:11, Sugiyama:20, Umegaki:17,Sugiyama:20,Ohishi:22a,Umegaki:22,Forslund:25}. This is orders of magnitude smaller than the value expected in the classical over-barrier-jump diffusion, as $\nu_0$ is expected to be within the range of the Debye frequency $\nu_{\rm D}=k\Theta_{\rm D}/h\approx10^{12}$--$10^{13}$~s$^{-1}$ regardless of diffusing species \cite{Mehrer:07, Shewmon:16}. No reasonable explanation for this huge discrepancy has yet been provided.

\begin{figure}[t]
 \begin{center}
 \includegraphics[scale=0.32,clip]{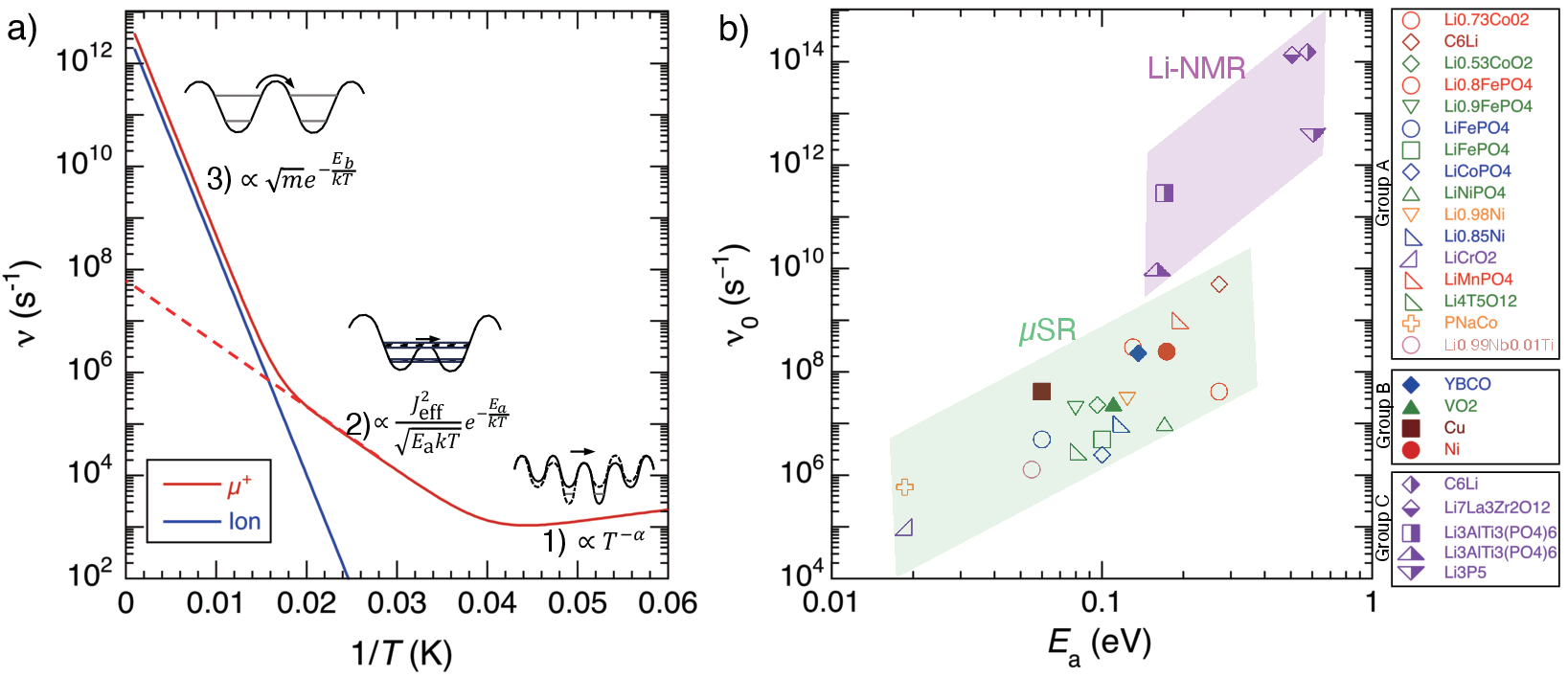}
 \caption{(a) Three modes of muon diffusion:
(1) coherent/incoherent tunneling ($T < 0.2$--$0.3\Theta_{\rm D}$),
(2) phonon-assisted tunneling ($T > 0.2$--$0.3\Theta_{\rm D}$),
and (3) classical over-barrier jump ($T > \Theta_{\rm D}$), where $\Theta_{\rm D}$ is the Debye temperature. Note that only the mode (3) is effective for ion diffusion.
(b) Activation energy $E_{\rm a}$ and attempt frequency $\nu_0$ extracted from $\mu$SR literature reporting ion diffusion (Group A: 
Li$_{0.73}$CoO$_2$, Li$_{0.73}$CoO$_2$\cite{Sugiyama:09}, 
C$_6$Li\cite{Umegaki:17}, 
Li$_{0.8}$FePO$_4$, Li$_{0.9}$FePO$_4$, LiFePO$_4$\cite{Baker:11},  
LiFePO$_4$, LiCoPO$_4$, LiNiPO$_4$\cite{Sugiyama:12}, 
Li$_{0.98}$NiO$_2$, Li$_{0.85}$NiO$_2$\cite{Mansson:13}, 
LiMnPO$_4$\cite{Sugiyama:20}, 
Li$_4$Ti$_5$O$_{12}$\cite{Umegaki:22}, 
P2-NaCoO$_2$\cite{Ohishi:23}, 
Li$_{0.99}$Nb$_{0.01}$TiO$_2$\cite{Forslund:25}), 
muon diffusion (Group B: YBCO\cite{Nishida:91}, VO$_2$\cite{Okabe:24}, Cu\cite{Kadono:89}, Ni\cite{Graf:79}), and Li-ion diffusion measured by $^7$Li-NMR (Group C: C$_6$Li\cite{Langer:13}, Li$_7$La$_3$Zr$_2$O$_{12}$\cite{Kuhn:11}, Li$_{1.5}$Al$_{0.5}$Ti$_{1.5}$(PO$_4$)$_3$\cite{Epp:15}, Li$_3$P$_5$O$_{14}$\cite{Duff:24}).}  \label{fig3}
 \end{center}
\end{figure}

Our re-evaluation of the earlier $\mu$SR results on Na$_x$CoO$_2$ demonstrates that the spin relaxation observed in $\mu^+$SR spectra is not primarily due to Na-ion diffusion, but is instead dominated by muon self-diffusion \cite{Ito:24,Ito:25b}. This new physical picture naturally resolves the long-standing puzzle of the unphysically small attempt frequency, as the observed dynamics are not governed by the classical diffusion of heavy sodium ions.  This finding calls for further review of the past $\mu$SR works on the ion dynamics based on the ``immobile-muon'' assumption. Although it is emphasized that the difference between hydrogen and muons is small enough to be ignored when discussing local electronic structures, it is important to note that the quantum properties of light muons cause significant differences from ions in dynamical properties such as diffusion.

More specifically, implanted muons form small polaron states with local lattice distortions in the host material, and they exhibit diffusive motion by three different mechanisms with increasing temperature (see Figure \ref{fig3}(a)): (1) incoherent/coherent tunneling to neighboring sites with lattice distortion where $\nu$ follows an inverse-power law in temperature, $T^{-\alpha}$, (2) phonon-assisted tunneling between degenerate states temporarily recovered by eliminating the lattice distortion upon phonon absorption, and (3) classical over-barrier jump \cite{Storchak:98,Fukai:05}. Of these, $\nu$ in (2) and (3) shows apparently the same Arrhenius-type temperature dependence, but their $\nu_0$ and $E_{\rm a}$ differ significantly between (2) and (3) due to the difference in the diffusion mechanism. Specifically, in (2), $\nu_0$ is given by $\nu_0= J_{\rm eff}^2/h\sqrt{E_{\rm a}kT}\approx10^6$--$10^9$ s$^{-1}$ with $J_{\rm eff}$ being the effective tunneling matrix determined by the overlap integral of the muon wave functions between degenerate states, and $E_{\rm a}$ is a fraction of the potential barrier $E_{\rm b}$ for the over-barrier jump \cite{Flynn:70}. Meanwhile, $\nu_0\approx\nu_{\rm D}$  and $E_{\rm a}=E_{\rm b}$  in (3).

Here, we extracted these two parameters from literature claiming to have measured {\sl ion} diffusion (Group A), and plotted them with $E_{\rm a}$ and $\nu_0$ respectively on the horizontal and vertical axes in Figure \ref{fig3}(b). For comparison, the values from papers reporting {\sl muon} diffusion (Group B) and those claiming to have measured Li ion diffusion by NMR (Group C) are superimposed. As can be seen, Groups A and B have in common smaller $\nu_0$ and $E_{\rm a}$ than those in Group C. Considering that diffusion by mechanisms (1) and (2) is negligible in ion diffusion, it is strongly suggested that the results for Group A are dominated by muon self-diffusion in common with Group B. Attention should be also paid to the cases where muon motion occurs by the mechanisms other than phonon-assisted tunneling; it can be driven by local tunneling or a significant lowering of the effective energy barrier due to pronounced zero-point oscillations \cite{Ito:23,Dehn:21,Hempelmann:98,Ito:17}. Meanwhile, the mechanism (3) also dominates in muon diffusion at higher temperatures, as has been demonstrated in several compounds \cite{Graf:79,Ito:10} (not shown here). 

As has been confirmed in the hybrid perovskite example, the temperature dependence of the $\mu$SR spectra for $Q<1$ is qualitatively different from that expected from the dKT function corresponding to $Q=1$.
Considering that the change in the $\mu$SR spectra for all Group A cases is well reproduced by the conventional  dKT function where the deduced $\nu$ follows the Arrhenius equation, it is more likely that the spin relaxation observed in these cases is in fact primarily due to muon self-diffusion rather than ion diffusion.

Finally, we emphasize that there are other complexities to be addressed for establishing $\mu$SR as a feasible technique for studying the ion dynamics. In particular, when a transition from a metastable muon state generated by electron excitations due to kinetic energy \cite{Okabe:24,Kadono:24} to a more stable state occurs within the $\mu$SR time window (which is often the case), a model that takes into account the accompanying change in $\nu$ will be necessary \cite{Ito:23,Hempelmann:98}. Furthermore, the electrostatic potential of $\mu^+$ can create correlated motions between the probe and surrounding ions. A complete understanding of such correlated, quantum-mechanical motion beyond the adiabatic approximation may require advanced simulation methods like Path Integral Molecular Dynamics (PIMD) and its derivative techniques \cite{Miyake:98, Kimizuka:18}.

In conclusion, our extended dKT function provides an essential first step to correctly interpret experimental data and to distinguish between the dynamics of the ion and the probe itself, paving the way toward a more reliable understanding of ion dynamics in solids.
\vspace{2mm}

This work was financially supported in part by JSPS Grants-in-Aid for Scientific Research (Grant No.~24H00477, 23K11707, 21H05102, and 20H01864), and by the MEXT Program: Data Creation and Utilization Type Material Research and Development Project (Grant No. JPMXP1122683430). 

%\bibliography{Refs}

\begin{thebibliography}{10}
\expandafter\ifx\csname url\endcsname\relax
  \def\url#1{{\tt #1}}\fi
\expandafter\ifx\csname urlprefix\endcsname\relax\def\urlprefix{URL }\fi
\providecommand{\eprint}[2][]{\url{#2}}

\bibitem{Sugiyama:09}
Sugiyama J, Mukai K, Ikedo Y, Nozaki H, M{\aa}nsson M and Watanabe I 2009 {\em
  Phys. Rev. Lett.\/} {\bf 103}(14) 147601

\bibitem{Sugiyama:11}
Sugiyama J, Nozaki H, Harada M, Kamazawa K, Ofer O, M\aa{}nsson M, Brewer J~H,
  Ansaldo E~J, Chow K~H, Ikedo Y, Miyake Y, Ohishi K, Watanabe I, Kobayashi G
  and Kanno R 2011 {\em Phys. Rev. B\/} {\bf 84}(5) 054430

\bibitem{Sugiyama:12}
Sugiyama J, Nozaki H, Harada M, Kamazawa K, Ikedo Y, Miyake Y, Ofer O,
  M\aa{}nsson M, Ansaldo E~J, Chow K~H, Kobayashi G and Kanno R 2012 {\em Phys.
  Rev. B\/} {\bf 85}(5) 054111

\bibitem{Mansson:13}
M{\aa}nsson M and Sugiyama J 2013 {\em Phys. Scr.\/} {\bf 88} 068509

\bibitem{Baker:11}
Baker P~J, Franke I, Pratt F~L, Lancaster T, Prabhakaran D, Hayes W and
  Blundell S~J 2011 {\em Phys. Rev. B\/} {\bf 84}(17) 174403

\bibitem{Umegaki:17}
Umegaki I, Kawauchi S, Sawada H, Nozaki H, Higuchi Y, Miwa K, Kondo Y,
  M{\aa}nsson M, Telling M, Coomer F~C, Cottrell S~P, Sasaki T, Kobayashi T and
  Sugiyama J 2017 {\em Phys. Chem. Chem. Phys.\/} {\bf 19}(29) 19058--19066

\bibitem{Sugiyama:20}
Sugiyama J, Umegaki I, Takeshita S, Sakurai H, Nishimura S, Forslund O~K,
  Nocerino E, Matsubara N, M\aa{}nsson M, Nakano T, Yamauchi I, Ninomiya K,
  Kubo M~K and Shimomura K 2020 {\em Phys. Rev. B\/} {\bf 102}(14) 144431

\bibitem{Ohishi:22a}
Ohishi K, Igarashi D, Tatara R, Nishimura S, Koda A, Komaba S and Sugiyama J
  2022 {\em ACS Phys. Chem Au\/} {\bf 2} 98--107

\bibitem{Ohishi:22b}
Ohishi K, Igarashi D, Tatara R, Umegaki I, Koda A, Komaba S and Sugiyama J 2022
  {\em ACS Appl. Energy Mater.\/} {\bf 5} 12538--12544

\bibitem{Umegaki:22}
Umegaki I, Ohishi K, Nakano T, Nishimura S, Takeshita S, Koda A, Ninomiya K,
  Kubo M~K and Sugiyama J 2022 {\em J. Phys. Chem. C\/} {\bf 126} 10506--10514

\bibitem{Ohishi:23}
Ohishi K, Igarashi D, Tatara R, Umegaki I, Nakamura J~G, Koda A, M{\aa}nsson M,
  Komaba S and Sugiyama J 2023 {\em ACS Appl. Energy Mater.\/} {\bf 6}
  8111--8119

\bibitem{Forslund:25}
Forslund O~K, Cavallo C, Cedervall J, Sugiyama J, Ohishi K, Koda A, Latini A,
  Matic A, M{\aa}nsson M and Sassa Y 2025 {\em Carbon Energy\/}  e70017

\bibitem{Ito:24}
Ito T~U and Kadono R 2024 {\em J. Phys. Soc. Jpn.\/} {\bf 93} 044602

\bibitem{edKT}
https://github.com/tuito0/musrfit-dynGssEALF (accessed on July 2025)

\bibitem{Edwards:75}
Edwards S~F and Anderson P~W 1975 {\em J. Phys. F: Metal Phys.\/} {\bf 5} 965

\bibitem{Edwards:76}
Edwards S~F and Anderson P~W 1976 {\em J. Phys. F: Metal Phys.\/} {\bf 6} 1927

\bibitem{Ito:25b}
Ito T~U and Kadono R 2025 {\em arXiv:2505.09147\/}

\bibitem{Hiraishi:23}
Hiraishi M, Koda A, Okabe H, Kadono R, Dagnall K~A, Choi J~J and Lee S~H 2023
  {\em J. Appl. Phys.\/} {\bf 134} 055106

\bibitem{Mehrer:07}
Mehrer H 2007 {\em Diffusion in Solids\/} (Springer Berlin, Heidelberg) ISBN
  978-3-540-71486-6

\bibitem{Shewmon:16}
Shewmon P 2016 {\em Diffusion in Solids\/} (Springer Cham) ISBN
  978-3-319-48564-5

\bibitem{Nishida:91}
Nishida N and Miyatake H 1991 {\em Hyperfine Interact.\/} {\bf 63} 183--197

\bibitem{Okabe:24}
Okabe H, Hiraishi M, Koda A, Matsushita Y, Ohsawa T, Ohashi N and Kadono R 2024
  {\em Phys. Rev. Mater.\/} {\bf 8}(2) 024602

\bibitem{Kadono:89}
Kadono R, Imazato J, Matsuzaki T, Nishiyama K, Nagamine K, Yamazaki T, Richter
  D and Welter J~M 1989 {\em Phys. Rev. B\/} {\bf 39}(1) 23--41

\bibitem{Graf:79}
Graf H, Holzschuh E, Recknagel E, Weidinger A and Wichert T 1979 {\em Hyperfine
  Interact.\/} {\bf 6} 245--249

\bibitem{Langer:13}
Langer J, Epp V, Heitjans P, Mautner F~A and Wilkening M 2013 {\em Phys. Rev.
  B\/} {\bf 88}(9) 094304

\bibitem{Kuhn:11}
Kuhn A, Narayanan S, Spencer L, Goward G, Thangadurai V and Wilkening M 2011
  {\em Phys. Rev. B\/} {\bf 83}(9) 094302

\bibitem{Epp:15}
Epp V, Ma Q, Hammer E~M, Tietz F and Wilkening M 2015 {\em Phys. Chem. Chem.
  Phys.\/} {\bf 17}(48) 32115--32121

\bibitem{Duff:24}
Duff B~B, Corti L, Turner B, Han G, Daniels L~M, Rosseinsky M~J and Blanc F
  2024 {\em Chemistry of Materials\/} {\bf 36} 7703--7718

\bibitem{Storchak:98}
Storchak V~G and Prokof'ev N~V 1998 {\em Rev. Mod. Phys.\/} {\bf 70}(3)
  929--978

\bibitem{Fukai:05}
Fukai Y 2005 {\em The Metal-Hydrogen System\/} (Springer Berlin, Heidelberg)

\bibitem{Flynn:70}
Flynn C~P and Stoneham A~M 1970 {\em Phys. Rev. B\/} {\bf 1}(10) 3966--3978

\bibitem{Ito:23}
Ito T~U, Higemoto W and Shimomura K 2023 {\em Phys. Rev. B\/} {\bf 108}(22)
  224301

\bibitem{Dehn:21}
Dehn M~H, Shenton J~K, Arseneau D~J, MacFarlane W~A, Morris G~D, Maign\'e A,
  Spaldin N~A and Kiefl R~F 2021 {\em Phys. Rev. Lett.\/} {\bf 126}(3) 037202

\bibitem{Hempelmann:98}
Hempelmann R, Soetratmo M, Hartmann O and W{\"a}ppling R 1998 {\em Solid State
  Ionics\/} {\bf 107} 269--280

\bibitem{Ito:17}
Ito T~U, Koda A, Shimomura K, Higemoto W, Matsuzaki T, Kobayashi Y and Kageyama
  H 2017 {\em Phys. Rev. B\/} {\bf 95}(2) 020301

\bibitem{Ito:10}
Ito T~U, Higemoto W, Ohishi K, Nishida N, Heffner R~H, Aoki Y, Suzuki H~S,
  Onimaru T, Tanida H and Takagi S 2010 {\em J. Phys.: Conf. Ser.\/} {\bf 225}
  012021

\bibitem{Kadono:24}
Kadono R and Hosono H 2024 {\em Adv. Phys.\/}  DOI:
  10.1080/00018732.2024.2413342

\bibitem{Miyake:98}
Miyake T, Ogitsu T and Tsuneyuki S 1998 {\em Phys. Rev. Lett.\/} {\bf 81}(9)
  1873--1876

\bibitem{Kimizuka:18}
Kimizuka H, Ogata S and Shiga M 2018 {\em Phys. Rev. B\/} {\bf 97}(1) 014102

\end{thebibliography}
%\setlength{\parskip}{1mm}
\providecommand{\newblock}{}

\end{document}